\documentstyle[aaspp4,epsfig,tighten,flushrt,multirow]{article}
\def\simlt{\lower.5ex\hbox{$\; \buildrel < \over \sim \;$}}
\def\simgt{\lower.5ex\hbox{$\; \buildrel > \over \sim \;$}}

\def\kms{\mbox{ km s$^{-1}$}}
\def\mpc{\mbox{ Mpc}}
\def\kpc{\mbox{ kpc}}

\def\msun{\mbox{ M}_\odot}
\def\mas{\mbox{ mas}}
\newcommand\beq{\begin{equation}}
\newcommand\eeq{\end{equation}}

\begin{document}

\title{The Detection of Pure Dark Matter Objects with Bent Multiply Imaged
Radio Jets}

\author{R. Benton Metcalf}
\affil{\it Institute of Astronomy, University of Cambridge, Cambridge CB3 0HA, UK} 

\begin{abstract}
When a gravitational lens produces two or more images of a quasar's
radio jet the images can be compared to reveal the presence of
small structures along one or more of the lines of sight.  If mass is
distributed smoothly on scales of $\simlt 10^7\msun$
independent bends in the jet images on milli-arcsecond 
scales will not be produced.
All three of the well collimated multiply imaged radio jets that have
been mapped on milli-arcsecond scales show some evidence of independent
bends in their images.  
Using existing data we model the lens system B1152+199 and show that it
likely contains a substructure of mass $\sim 10^5 - 10^7h^{-1}\msun$ or
a velocity dispersion of $\sim 10\kms$.  An alternative explanation is that an
intrinsic bend in the jet is undetected in one image and magnified in the
other.  This explanation is disfavored and future observations could
remove any ambiguity that remains.
The  probability of a radio jet being bent by small scale
structure both inside and outside of the host lens is then investigate.
The known populations of dwarf galaxies and globular clusters
are far too small to make this probability acceptable.  A
previously unknown population of massive dark objects is needed.  The
standard Cold Dark Matter (CDM) model might be able to account for the
observations if small mass halos are sufficiently compact.  In other
cosmological models where small scale structure is suppressed, such as
standard Warm Dark Matter (WDM), the observed bent jets would be very
unlikely to occur.
\end{abstract}

\section{Introduction}
\label{sec:introduction}

The standard $\Lambda$CDM cosmological model has been very
successful in accounting for observations on scales larger than  
around a Mpc.  However, it appears that this model faces 
difficulties on the scales of galaxies and dwarf galaxies
\markcite{2000AJ....119.1579V}({van den Bosch} {et~al.} 2000).  One such problem is that CDM
simulations of the local group of galaxies predict an order of magnitude
more dwarf galaxy halos with masses greater than $\sim 10^7\,\msun$ than
there are observed satellites of the Milky Way (MW) Galaxy and M31
\markcite{1999ApJ...524L..19M,1999ApJ...522...82K,1998ARA&A..36..435M}({Moore} {et~al.} 1999; {Klypin} {et~al.} 1999; {Mateo} 1998). 
These simulations predict that 10-15\% of the virial mass of a galaxy
halo is in substructures of mass $\simgt 10^7\msun$.

This over prediction of dwarf halos could be a sign that there is
something fundamentally wrong with the CDM model.  Proposed explanations
include warm dark matter (WDM) which smoothes out small scale structure in the
early universe \markcite{2001ApJ...556...93B}(e.g. {Bode}, {Ostriker}, \&  {Turok} 2001), unorthodox inflation models
which break scale invariance \markcite{2000PRL.Kamionkowski}(Kamionkowski \& Liddle 2000) and
self-interacting dark matter which causes substructures to evaporate
within larger halos \markcite{2000PhRvL..84.3760S}({Spergel} \& {Steinhardt} 2000).
Alternatively, CDM could be correct and the small Dark Matter (DM) clumps 
could exist, but not contain stars, so as to escape detection as
observable dwarf galaxies.  This situation can easily, perhaps
inevitably, come about through the action of feedback processes
(radiation and supernova winds) from the first generation of stars
in the universe \markcite{2000ApJ...539..517B,astro-ph/0107507}e.g. {Bullock}, {Kravtsov}, \&  {Weinberg} (2000); Somerville (2002).
For example, photoionization can prevent gas from
cooling and thus inhibit star formation in halos that are too small to
be self--shielding.  Several authors, \markcite{astro-ph/0109347}(e.g. Metcalf 2001),
have argued that the overabundance of DM clumps is likely to extend down
to smaller masses and larger fractions of the halo mass than have thus
far been accessible to numerical simulations.
These nearly pure dark matter structures have largely been considered
undetectable.

Gravitational microlensing by stars has been observed in the 4--image system
Q2237+0305 through the long term variations of the optical flux ratios \markcite{1989AJ.....98.1989I,witt:95,2000ApJ...529...88W}{Irwin} {et~al.} (1989); {Witt}, {Mao}, \& {Schechter} (1995); {Wo{\' z}niak} {et~al.} (2000, and references there in).
\markcite{1998MNRAS.295..587M}{Mao} \& {Schneider} (1998) first proposed larger scale substructure as an explanation
for the magnification ratios of the 4--image quasar lenses B1422+231
which do not agree with any simple lens model.  The modeling of
B1422+231 has since been improved in \markcite{astro-ph/0112038}Brada\v{c} {et~al.} (2002) and
\markcite{Keeton2002}Keeton (2002).  It still 
appears that a substructure with a mass of $10^4 - 10^7\,h^{-1}\msun$
near image~A is required to explain the difference between the radio and
optical flux ratios in this system.
\markcite{2001ApJ...563....9M}{Metcalf} \& {Madau} (2001) showed that if CDM substructure exists it
could be detected through the magnification ratios of 4--image quasar
lenses.  Concurrently \markcite{2002ApJ...565...17C}{Chiba} (2002) modeled three 4--image
lenses and showed that a significant amount of substructure was
necessary to make their magnification ratios agree with simple smooth
lens models.  These ideas have been further investigated in
\markcite{2002ApJ...567L...5M}{Metcalf} \& {Zhao} (2002) and \markcite{Dalal2002}Dalal \& Kochanek (2002).
These studies all rely on the influence of substructure on magnification
ratios.  This is a promising approach, but it is strongly model
dependent and susceptible to misinterpretation because of 
microlensing by ordinary stars in the lens galaxy.

It was also predicted in \markcite{2001ApJ...563....9M}{Metcalf} \& {Madau} (2001) that CDM
substructure should occasionally distort multiply imaged radio jets on
milli-arcsecond scales.  This distortion would not be reproduced in all
the images so it can be distinguished from structure in the jet itself.
This effect had also been suggested by
\markcite{1992ApJ...397L...1W}{Wambsganss} \& {Paczynski} (1992) as a method for detecting a large abundance
of $m \simgt 10^6\msun$ primordial black holes.  Previous to this
\markcite{1981ApJ...246....1B}{Blandford} \& {Jaroszynski} (1981) had considered the distortion of singly imaged radio
jets as a probe of galaxies under the assumption that they are
intrinsically straight.  As will be demonstrated, the method
considered here has the important advantages over magnification ratios methods
of avoiding any confusion with microlensing and avoiding any strong
dependence on the lens model.

In section~\ref{sec:observ-b1152+199} the observations of mapped multiply
imaged radio jets are summarized.
In section~\ref{sec:modeling-jet} general considerations related to
modeling multiply imaged radio jets are discussed and specific models
for one particular case are presented.  The interpretation of these results in
terms of the level of small scale structure in the universe is addressed
in \S~\ref{sec:impl-dark-matt}.  General discussion and conclusions are
in \S~\ref{sec:discussion}.

In this paper the Hubble parameter is $H_0=65\, h_{65} \kms \mpc^{-1}$.
\footnote{On a couple of occasions when quoting other peoples work the convention
$H_0=100\, h \kms \mpc^{-1}$ is used.}  For quantities that do not have a
simple dependence on $H_o$ a value $h_{65}=1$ is used.  The
present average density of matter in the universe in units of the
critical density is $\Omega_m$ and the 
cosmological constant in the same units is $\Omega_\Lambda$.  The
``concordance'' cosmological model ($\Omega_m=0.3$, $\Omega_\Lambda=0.7$)
will be assumed throughout.  Milli-arcseconds will be abbreviated as mas.

\section{Observations of multiply imaged radio jets}
\label{sec:observ-b1152+199}

Several lensed QSO radio jets have been imaged on milli-arcsecond scales
with the Very Long Baseline Array (VLBA) and other Very Long Baseline
Interferometer (VLBI) configurations 
\markcite{1994MNRAS.270..457G,1997MNRAS.289..450K,1999MNRAS.303..727K,2001AJ....122..591R,2000evn..proc...49X,2000A&A...362..845R,2001ApJ...562..649K,2001AJ....121..619M,2002MNRAS.330..205R}({Garrett} {et~al.} 1994; {King} {et~al.} 1997; {Koopmans} {et~al.} 1999; {Rusin} {et~al.} 2001; {Xanthopoulos} {et~al.} 2000; {Ros} {et~al.} 2000; {Kemball}, {Patnaik}, \&  {Porcas} 2001; {Marlow} {et~al.} 2001; {Rusin} {et~al.} 2002).
In only three of these cases is the jet collimated enough and the resolution high
enough that a bend or kink could in principle be detected.  
 
The two image gravitational lens B1152+199 was discovered in the CLASS
radio survey and follow--up observations were done on the Keck II
telescope \markcite{1999AJ....117.2565M}({Myers} {et~al.} 1999).  The images are
separated by $1''.56$ and the redshifts of the source and lens are
$z_s=1.019$ and $z_l=0.439$.   Subsequently,
\markcite{2002MNRAS.330..205R}{Rusin} {et~al.} (2002) observed B1152+199 using the Hubble Space
Telescope (HST), the Multi-Element Radio-Linked Interferometer Network
(MERLIN) and VLBA.  In the HST observations a faint, indistinct lens
galaxy can be seen along with a fainter object which is interpreted as a
dwarf galaxy companion.  With VLBI they were able to map the two images of
the radio jet on milli--arcsecond scales.  They discovered that in
image~A the jet appears straight while in image~B it is bent.  
No formal constraint on the significants of this bend are given in
\markcite{2002MNRAS.330..205R}{Rusin} {et~al.} (2002) and further observations may be required to
make the detection certain.  For the purposes of this paper we will take
the observations at face value and assume the bend is not an instrumental
effect.
In section~\ref{sec:results-b1152+199} lensing
explanations for this bend are investigated .  The bend is clearly not
aligned with either the direction to image~A or to the lens galaxy.
Superluminal motion is a 
possible explanation only if the jet's shape can change on a
time scale that is smaller than the time delay between images.
\markcite{2002MNRAS.330..205R}{Rusin} {et~al.} (2002) fit a variety of smooth models to the
macroscopic lens and get time delays of 41.1 to 70.6~$h_{65}^{-1}$~days which
making this an unlikely explanation.  They do not attempt to explain the
bend with their lens models.

The four image lens MG J0414+0534 was observed with global VLBI by
\markcite{2000A&A...362..845R}{Ros} {et~al.} (2000).  The jet consists of a two component core and
two radio lobes on either side.  In images~A2 and B all the
radio components are nearly collinear while in image~A1 they are
drastically misaligned.  Only two components are detected in image~C so in
this case the alignment cannot be determined.  The
distortion of image~A1 could be caused by a substructure near the image
or it might be due to the magnification of a small misalignment in the
other images (see section~\ref{sec:with-no-substructure}).  The
situation will be clarified with further modeling of this particular source.

The double quasar Q0957+561 was the first gravitational lens discovered
\markcite{1979Natur.279..381W}({Walsh}, {Carswell}, \&  {Weymann} 1979) and has been studied extensively in the past
two decades.  The VLBI maps of the radio jets
 appear to show a kink in image~A that is not
reproduced in image~B (near $\Delta\delta = 20$~mas, $\Delta\alpha = 10$~mas
with respect to the core)
\markcite{1994MNRAS.270..457G,1999ApJ...520..479B}({Garrett} {et~al.} 1994; {Barkana} {et~al.} 1999).  Although in this case
the bend is much less certain than in B1152+199 or MG J0414+0534-- and we will not try to
reproduce it with a lens model here -- it does suggest that milli-arcsecond
kinks and bends are common.  This has very important consequences in
relation to the discussion in \S~\ref{sec:impl-dark-matt}, because it
implies that the bend in B1152+199 is not just a rare coincidental
alignment of the image and a known type of substructure.

\section{Modeling the Jet}
\label{sec:modeling-jet}

\subsection{Formalism}
\label{sec:formalism}

The radio jet will be treated as a one dimensional curve on the sky
described by $\vec{\theta}_{\rm source}(s)$ in the absence of 
lensing.  An image of the jet is described by $\vec{\theta}_{\rm image}(s)$. 
The curve of the source jet is related to the curve of its image
through the lensing equation
\begin{eqnarray}\label{lens_eq}
{\bf y}(s) =  {\bf x}(s) -  {\bf \nabla}\psi \left({\bf x}(s) \right)
\end{eqnarray}
\begin{eqnarray}
{\bf y}(s)\equiv D_l \vec{\theta}_{\rm source}(s) / \lambda_o 
~~~~~~{\bf x}(s)\equiv D_l \vec{\theta}_{\rm image}(s) /\lambda_o 
\end{eqnarray}
where $\lambda_o$ is an arbitrary scaling length and $s$ is the
arc-length along the jet in the image plane measured in the same units
as ${\bf x}$.  The angular size distances to the lens, source, and from
the lens to the source will be denoted $D_l$, $D_s$, and $D_{\rm ls}$
respectively.  The lensing potential is related to the lens surface
density, $\Sigma({\bf x})$, through the Poisson equation
$\nabla^2\psi({\bf x}) = 2\kappa({\bf x})$ where $\kappa\equiv
\Sigma({\bf x})/\Sigma_c$.  The critical surface density is defined as
$\Sigma_c=(4\pi G D_lD_{ls}/c^2 D_s)^{-1}$.

The tangent and normal vectors of the jet are given by
\begin{eqnarray}
{\bf t}(s) \equiv \frac{\partial {\bf x}}{\partial s} ~~~~~~~~~~~~~~~~~~~~
{\bf n}(s) \equiv \frac{\partial^2 {\bf x}}{\partial s^2}.
\end{eqnarray}
The magnitudes of these vectors are $t(s)=1$ and $n(s)=1/R(s)$ where
$R(s)$ is the radius of curvature.  For convenience we define the
matrices 
\begin{eqnarray}\label{def:AB}
{\bf A}_{ij} \equiv  \delta_{ij} - \frac{\partial^2 \psi}{\partial x^i\partial x^j} ~~~~~~~~~~
{\bf M}_{ijk} \equiv -\frac{\partial^3 \psi}{\partial x^i\partial x^j\partial x^k}.
\end{eqnarray}
Now we can find the curvature and normal vectors to the source jet by
taking derivatives of the lens equation
\begin{eqnarray}\label{tangent}
{\bf T}(s) \equiv \frac{\partial {\bf y}}{\partial s'} = \frac{\partial
s}{\partial s'} \frac{\partial {\bf y}}{\partial s} = \frac{ {\bf u} }{
|{\bf u} | }
\end{eqnarray}
\begin{eqnarray}\label{Ncurv}
{\bf N}(s) \equiv \frac{\partial^2 {\bf y}}{\partial s'^2} =
\left(\frac{\partial s}{\partial s'}\right)^2 \frac{\partial^2 {\bf
y}}{\partial s^2} + \frac{\partial^2 s}{\partial s'^2} \frac{\partial {\bf
y}}{\partial s}
= \frac{1}{|{\bf u}|^{2}}\left( {\bf v}  - \frac{ {\bf u}( {\bf v}\cdot{\bf
u})}{|{\bf u}|^2} \right)  ,
\end{eqnarray}
\begin{eqnarray}\label{uandv}
{\bf u}_i\equiv\sum_j {\bf A}_{ij}{\bf t}_j~~~~~~~~{\bf v}_i\equiv\sum_j
{\bf A}_{ij}{\bf n}_j
+ \sum_{jk} {\bf M}_{ijk}{\bf t}_j{\bf t}_k
\end{eqnarray}
where $s'$ is the arc-length on the source plane.  The vectors ${\bf
T}(s)$ and ${\bf N}(s)$ must be the same for all images of the jet so
they can be used as constraints on the lens model.  Along with the position
coordinates on the source plane 
this makes 4 constraints per point on the jet (${\bf T}(s)$ and ${\bf
N}(s)$ must be perpendicular and $|{\bf T}(s)|=1$).
 
Let us estimate the relative size of the terms in~(\ref{Ncurv}).
For any spherically 
symmetric lens the Einstein ring radius, $\lambda_E$, is the solution to
\begin{equation}
\lambda_E^2 = \frac{M(\lambda_E)}{\pi \Sigma_c}
\end{equation}
where $M(\lambda_E)$ is the mass within a projected distance of
$\lambda_E$.  
Images that are significantly magnified form near the
Einstein radius for a spherical lens or, more generally, near critical
curves (the curve ${\bf x}$ where $\det[{\bf A}({\bf x})] =0$).  The
magnitude of the deflection angle near $\lambda_E$ is $\alpha(x) \sim
\lambda_E/\lambda_o$ so if an image is formed both near the Einstein
radius of a host halo and near the Einstein radius of a subclump
their contributions to the deflection will differ by a factor of $\sim
\lambda_E^{\rm clump}/\lambda_E^{\rm host} \sim
(\sigma_{\rm clump}/\sigma_{\rm host})^2$.  The matrices~(\ref{def:AB})
involve further derivatives of the lensing potential so
that at the same point the two contributions to ${\bf A}_{ij}({\bf x})$
will be roughly equivalent while the contribution to ${\bf M}_{ijk}$
from the subclump will be larger than the 
host's by a factor of $\sim \lambda_E^{\rm host}/ \lambda_E^{\rm clump}
\sim (\sigma_{\rm host}/\sigma_{\rm clump})^2 \sim M_{\rm host}/m_{\rm
sub}$.  For dwarf galaxy sized substructures this is $\sim 100-10,000$. 
From equation~(\ref{Ncurv}) we see that to generate a curvature radius
of order the jet size, $\lambda_s$, $\lambda_s M_{ijk}/(\lambda_o
|u|)$ needs to be $\simgt 1$.  Roughly speaking only objects with
Einstein radii of order the source size can create a noticeable bend.

As a working definition we will say that substructure is present in the
lens when the bending matrix ${\bf M}$ in equation~(\ref{Ncurv}) is
important.  The effect of a smooth lens on the shape of a small source can
then be describe by the magnification matrix ${\bf A}$ alone.  This
definition will clearly be dependent on the size of the source and the
resolution of the observations.

If we believe that a significant gravitational bending of a jet is rare
enough that it is unlikely to happen to both of a pair of images (at
least at the same point on the jet) then the equations can be
significantly simplified.  The image without substructure will be
labeled image~2.  By expanding the lensing
equation~(\ref{lens_eq}) around a point on image~1 and the corresponding
point on image~2 and equating the position on the source plane we can
arrive at an equation analogous to~(\ref{Ncurv}) but relating the
curvature of one jet image to the curvature of the other:
\begin{eqnarray}\label{n2(n1)}
{\bf n}^{(2)} = \frac{1}{|{\bf \tilde{u}}|^{2}}\left( {\bf \tilde{v}}  -
\frac{ {\bf \tilde{u}}( {\bf \tilde{v}}\cdot{\bf \tilde{u}})}{|{\bf
\tilde{u}}|^2} \right).
\end{eqnarray}
The tildes signify the quantities in~(\ref{uandv}) only with the
matrices ${\bf\tilde{A}} = {\bf A_{(2)}^{-1}}{\bf A_{(1)}}$ and
${\bf\tilde{M}} = {\bf A_{(2)}^{-1}} {\bf M_{(1)}}$ substituted.

We can see from (\ref{n2(n1)}) that in the absence of substructure a jet
that is straight in one of its images will also be straight in its other
images.  However, the intrinsic curvature of the jet can be magnified or
demagnified without substructure.  In some cases the curvature
could be observed in one image, but be too small in another image to be
detected resulting in the erroneous conclusion that substructure must be
present.  For this reason it is important to quantify by how much the
curvature can be changed without substructure.  In this case
${\bf\tilde{M}}=0$ and from (\ref{n2(n1)}) we can find the curvature
magnification factor
\begin{equation}\label{curv_mag}
C\equiv \frac{|{\bf n^{(2)}}|}{|{\bf n^{(1)}}|} = \frac{1}{|{\bf\tilde{u}}|^2} \left[
|{\bf\tilde{v}^*}|^2 -
\frac{({\bf\tilde{v}^*}\cdot {\bf\tilde{u}})^2}{|{\bf\tilde{u}}|^2}
\right]^{1/2}
\end{equation}
where ${\bf\tilde{v}^*}$ is ${\bf\tilde{v}}$ with ${\bf n^{(1)}}$ replaced
with the unit vector ${\bf\hat{n}^{(1)}}$.  This quantity can be
calculated with a smooth lens model fit to the image positions and the
tangent vectors to the jet images.
\begin{figure}[t]
\centering\epsfig{figure=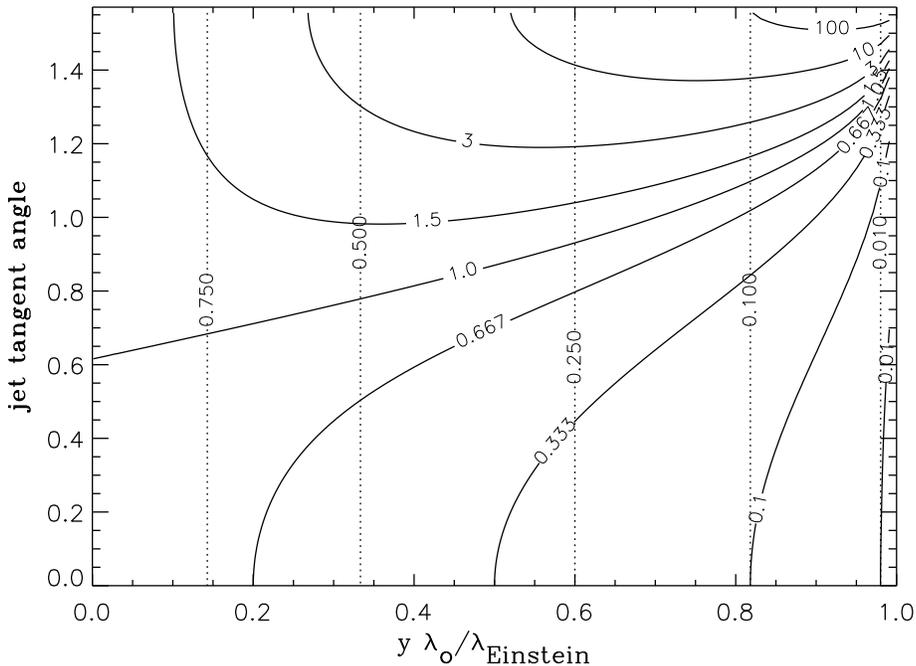,height=3.75in}
\caption[qiw]{\footnotesize The solid contours are curvature magnification
factor~(\ref{curv_mag}) for a singular isothermal sphere as a function
of the radial position of the source in Einstein ring radii and the
tangent vector to the outer jet image.  The source position $y$ is the
magnitude of the vector defined in equation~(\ref{lens_eq}).  The
tangent angle is defined so that zero is a radial jet and $\pi/2$ is
tangentially oriented.  For $C>1$ the inner image is more curved and for
$C<1$ the opposite is true.  For $y\lambda_o>\lambda_E$ there is only
one image.  The dotted contours show absolute value of the magnification
ratio of the inner image to the outer image.
}
\label{fig:curv_mag}
\end{figure}

It is useful to have concrete models for the lenses.
For a spherically symmetric lens with a power--law mass profile
($M(r)\propto r^n$), or at least a power--law near the location of the
image, the matrices~(\ref{def:AB}) can be calculated directly: 
\begin{equation}
{\bf \alpha}({\bf \tilde{x}})=\frac{\lambda_E}{\lambda_o}
\frac{{\bf \tilde{x}}}{\tilde{x}^{2-n}}
\end{equation}
\begin{equation}
A_{ij}=  \delta_{ij}-\frac{1}{\tilde{x}^{2-n}} \left[ \delta_{ij} - (2-n)
\frac{\tilde{x}^i \tilde{x}^j}{\tilde{x}^2} \right]
\end{equation}
\begin{equation}
M_{ijk}=\left( \frac{\lambda_o}{\lambda_E}
\right)\frac{(2-n)}{\tilde{x}^{3-n}} \left[ \frac{\delta_{ij}\tilde{x}^k + \delta_{ik}\tilde{x}^j+ \delta_{jk}\tilde{x}^i}{\tilde{x}} - (4-n) \frac{\tilde{x}^i \tilde{x}^j \tilde{x}^k}{\tilde{x}^3}
 \right]
\end{equation}
where ${\bf \tilde{x}} \equiv({\bf x} - {\bf x}_o) \lambda_o/
\lambda_E$ is the image position relative to the center of the lens.
Also useful is the convergence or dimensionless
surface density at the Einstein radius in these power--law model:
$\kappa(\lambda_E)=n/2$.
For a Singular Isothermal Sphere (SIS) lens $n=1$ and
\begin{equation}
\lambda_E=4\pi \left(\frac{\sigma}{c}\right)^2 \frac{D_l D_{ls}}{D_s} ~~~~,~~~
\rho(r)=\frac{\sigma^2}{2\pi G r^2}.
\end{equation}
For a point mass $n=0$ and $\lambda_E=\sqrt{m/(\pi\Sigma_c)}$.  

As an example, figure~\ref{fig:curv_mag} shows the curvature
magnification factor, $C$, for a SIS lens with no substructure.  This
quantity depends on both the position of the source and the tangent
vector to one of the images (in this case the outer image is chosen).
The factor is generally larger for jets that are radial (in which case
the outer image is more bent) or tangential (where the opposite is
true).  Cases with $C$ much different from one tend to have smaller
magnification factors in the sense that the outer image is much
brighter.  Because of this there will be a bias toward cases where $C$
is near one.  To fit real lens system a more complicated, asymmetric lens
models must be used and $C$ must be calculated for each pair of images
separately.  This quantity can be evaluated at the center of a jet
image or at a kink in a jet image to determine if the bend is consistent
with an intrinsic feature in the jet itself or requires substructure as an explanation.

\subsection{Modeling of B1152+199}
\label{sec:results-b1152+199}

Two explanations for the apparent bend in image~B of B1152+199 will be
explored.  One is that image~A actually has a small undetected curvature
which is magnified in image~B where it is detected.  The second explanation is 
that image~A is straight and image~B is bent by the influence of a
substructure near it.  Investigating both of these hypothesis
requires fitting a host lens model to the positions of the images and
the center of the lens.  Since there are only two images in this case a
complicated host lens model is not well constrained by the positions alone.
\markcite{2002MNRAS.330..205R}{Rusin} {et~al.} (2002) fit to each VLBI image a point source for the
core and a Gaussian for the jet; these positions are used as constraints.  
We choose to use a simple SIS model with a background shear -
$\alpha^1({\bf x}) = \gamma \left[ x^1\cos(2\theta_\gamma) +
x^2\sin(2\theta_\gamma)\right]$, $\alpha^2({\bf x}) = \gamma \left[
x^1\sin(2\theta_\gamma) - x^2\cos(2\theta_\gamma) \right]$.  The shear
breaks the azimuthal symmetry of the host lens which is necessary for it
to fit the observed lens position.  No attempt is made to incorporate
the possible dwarf companion of the lens galaxy that appears as a very
faint smudge in the HST image.  We do not expect that this object is
large enough to 
significantly change the surface potential except in its near vicinity
and the images are well separated from it.  In addition, the quality of
the fit discussed in \S~\ref{sec:with-no-substructure}
gives us confidence that the model accurately
reproduces the local magnification matrix at the positions of the
images which is the only thing needed here.  With the reported redshifts the
critical density for this lens is $\Sigma_c = 2.65\times 10^9
h_{65}\msun\kpc^{-2}$.

\subsubsection{no substructure}
\label{sec:with-no-substructure}
\begin{figure}[t]
\centering\epsfig{figure=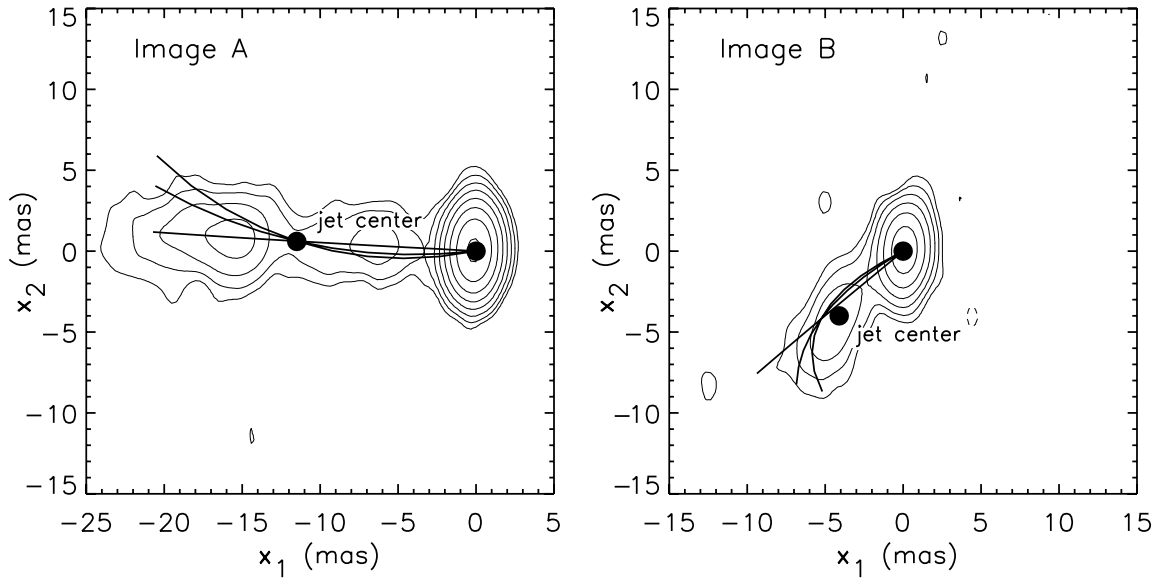,height=3.0in}
\caption[qiw]{\footnotesize Models for the jet shape with no
substructure.  Using the best--fit smooth model the arc in image~A is
mapped onto the B image.  The radio core is at the origin in both cases
and the jet center is marked.  The three arcs have curvatures of $|{\bf
n_A}| = 0$, 0.029 and 0.043 mas$^{-1}$ in image~A and are designed to
pass through both the core and the jet center.  The model curves are
overlayed on the \markcite{2002MNRAS.330..205R}{Rusin} {et~al.} (2002) VLBI map.  The lowest
contour is 3 times the rms noise and each contour is increased by a
factor of 2.  The beam is $3.6\times 1.9$ with the larger axes being in the vertical $x_2$ direction.
}
\label{fig:map_jets_nosub}
\end{figure}

A smooth model is fit to the positions of the lens galaxy, the radio cores of
the images and the center of the jet images.  A model is found that fits
all the positions to better than 0.1 milli-arcsecond.  In addition, the
magnification ratio of the radio core agrees with the observed one to
better than 10\% despite this not being used as a constraint on the
model.  This signifies that the local magnification matrix,
${\bf\tilde{A}}$, is being accurately reproduced by the model.
The velocity
dispersion of the lens is $\sigma_{\rm host}=247\kms$ and the
background shear is $\gamma=0.102$.    This velocity dispersion is not
unusual for a lens galaxy.  The estimated circular velocity is $V_{\rm
circ}=\sqrt{2}\sigma_{\rm host}$.  The magnifications at the positions of
the radio cores are $\mu_A=3.8$ and $\mu_B=-1.5$ -- a negative
magnification indicates a one dimensional parity flip in the image.
This model gives a curvature magnification factor of $C=4.9$ at the
center of the jet with image~B being the more curved of the two images
as observed.  If the jet in image~A has a curvature of $1/C$ times the curvature in image~B and it is in the right
direction then the observations can be explained without substructure.
Figure~\ref{fig:map_jets_nosub} shows some attempts to model the jet in
this way.  From visual inspection it appears that the jet in image~A is
not bent enough to explain the bend in image~B.  The curve should follow
the crest of the jet's surface brightness, but a jet that is
bent enough requires the end of the jet to be shifted by $\sim 3-4$~mas
from the crest of the straight jet.  The beam is large in this dimension,
$3.6$~mas, but a shift in the crest should be detectable below this
level.  

Another way of evaluating this is to realize that
\begin{equation}
C\simeq \left( \frac{\theta^A_{\rm jet}}{\theta^B_{\rm jet}} \right)^2
\frac{\delta^B}{\delta^A}
\end{equation}
where $\theta_{\rm jet}$ is the length of the jet image and
$\delta$ if the maximum deviation of the crest from a straight line.
Judging from \markcite{2002MNRAS.330..205R}{Rusin} {et~al.} (2002) $\theta^B_{\rm jet}\simeq
10\mas$, $\theta^A_{\rm jet}\simeq 22.0\mas$ and $\delta^B \simeq 2\mas$
giving $\delta^A \simeq 2\mas$ with the derived curvature magnification
factor.  This is small, but appears to be outside
of the 12$\sigma$ contours along the full length of the jet in their
map (the peaks in the jet are above 24$\sigma$).  A more conclusive determination will probably require improved
observations.
\subsubsection{substructure}
\label{sec:with-substructure}
\begin{figure}[t]
\centering\epsfig{figure=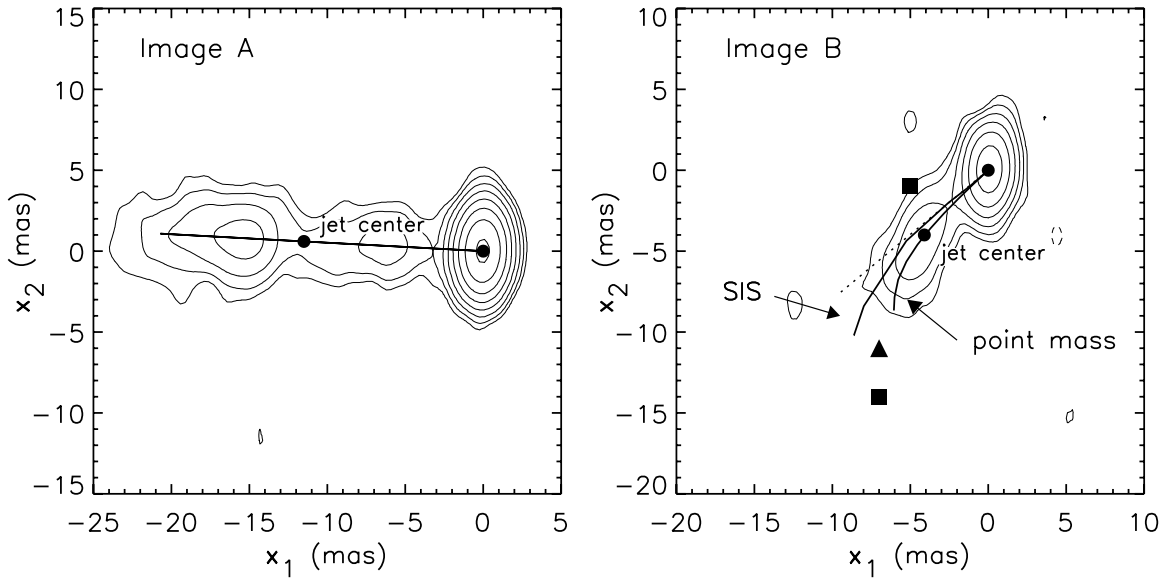,height=3.0in}
\caption[qiw]{\footnotesize These are diagrams showing the
reconstruction of Image~B from a line segment representing Image~A using the two
lens models discussed in the text.  The positions of the point mass
substructure is marked by the triangle and the positions of the SIS
substructures are marked by squares.  The jets corresponding to each
substructure model are marked with arrows.  The dotted
curve in the right hand panel is the image without
any substructure.  In the background is the \markcite{2002MNRAS.330..205R}{Rusin} {et~al.} (2002)
VLBI maps.
}
\label{fig:map_jets}
\end{figure}

The substructure is modeled by adding either SIS or point masses to the
smooth model described above.  Several different methods for fitting the
jet shape were tried.  An essential difficulty is that besides the core
there are no clear localized features along the jet that can be identified in
both images.  The positions of these features along with the tangent and
curvature at such points could have been used as constraints were they present.
Another difficulty arises from the large number of local minima in any
$\chi^2$ function that was tried -- there are different ways of bending a
straight image by either ``push'' or ``pulling'' at different points.  

It was found that the best and most unambiguous results were obtained by
first fixing the smooth, or host, lens model to the one discussed in
\S~\ref{sec:with-no-substructure}.  A straight line representing the jet
in image~A is then mapped to image~B using the model.  The substructures
are added near image~B by trial and error assisted by minimizing a
$\chi^2$ based on the positions of the core and jet center until a curve
in image~B is obtained that best reproduces the qualitative features of
the VLBI map.  This method does not use the observed
magnification ratio of the cores as a constraint so any possible
contamination from microlensing by stars is entirely avoided.
Figure~\ref{fig:map_jets} shows the results of this fitting. 
The resulting lens model is not unique in any quantitative sense, but there
are clear things that can be learned from this fitting process about the
kind of substructure that is capable of producing the bend.

When a point mass is used as a substructure the shape of image~B is
comparatively easy to reproduce.  A point mass can be considered an
approximation to any substructure that is very compact relative to its
own Einstein radius such as a tidally truncated dark matter halo.  Such a substructure can cause a strong deflection
near its center while having a limited range of influence.  This enables the
point mass substructure in figure~\ref{fig:map_jets} to displace the
lower end of the jet while leaving the position of the core end of the jet
relatively unchanged.  Note that the substructure has the effect of
attracting the image rather than repelling it as would normally be the
case.  This attraction happens in only one dimension and is a result of
one of the eigenvalues of the magnification matrix derived from the host
lens being negative (image~B is reflected in one dimension with respect
to image~A).  The mass is most naturally calculated in units of the mass
of the host lens within its Einstein radius which 
in this case is $M_E = (\sigma/c)^4 G^{-1}\Sigma_c(z_s,z_l)^{-1} =
1.6\times 10^{11} h_{65}^{-1} (\sigma/247\kms)^4 \msun$.  The favored
model has a substructure mass of $m=2.5\times 10^{-5} M_E$.  Other model
parameters are summarized in table~\ref{table}.  A point
mass with mass much more than $10^{-4} M_E$ tends to displace the lens
without creating a bend and a mass of $\simlt 10^{-6} M_E$ cannot
produce a bend on a large enough angular scale. 
\begin{table}[t]
\begin{tabular}{lccccll}
& \multicolumn{6}{c}{\bf Substructure Model Parameters}  \\
\multicolumn{1}{c}{} 
& $m/M_E$ & $\sigma (\kms)$ 
& $x^1$ (mas) & $x^2$ (mas) & $\kappa$ & $\mu_{\rm core}$ \\
Point mass  & $2.5\times 10^{-5}$ & - & -7.0 & -11.0 &
0.86 & -1.19 \\
\\
\multirow{2}{1cm}{SIS} & - & 9.6 & -5.0 & -1.0 &
\multirow{2}{1cm}{1.05} & \multirow{2}{1cm}{-1.07} \\
        & - & 21.0 & -7.0 & -14.0  \\
\end{tabular}
\caption{\footnotesize  The positions $x^1$ and $x^2$ are the center of
the substructure with respect to the core in image~B.  The surface
density $\kappa$ and the magnification $\mu_{\rm core}$ are evaluated at the core
in image~B. 
\label{table}}
\end{table}

When a SIS model is used for the substructure it is difficult to
reproduce the jet shape.  The tendency is that when the SIS is massive
enough to displace the lower end of the jet sufficiently it also displaces the
core end of the jet so that a significant bend is not created -- i.e. the SIS
model is not compact enough.  We partially get around this problem by
using two SIS substructures in figure~\ref{fig:map_jets} and
table~\ref{table}, but even this does not produce very satisfactory
results and considering the discussions in
section~\ref{sec:impl-dark-matt} this seems an improbable 
explanation.  It is possible that if the host lens model were allowed to
vary along with the sublens model an explanation could be found that
requires only one SIS substructure.  However, including the host lens in
the fitting process greatly increases 
the number of local minima in $\chi^2$ and after significant
experimentation we have been unable to find a model that is a
qualitative improvement on the one in figure~\ref{fig:map_jets} using a
single SIS substructure.  The two SIS model requires a precarious
balance between the effects of two relatively massive substructures.  A small
change in the positions or masses causes the jet to be rather drastically
distorted.  We conclude that substructures as diffuse as SISs are an
unlikely explanation for the observations.

By modeling the lens general conclusions can be made, but the specific
form of the substructure is not tightly constrained.
This modeling demonstrates that the bend in Q0957+561 can be reproduced
by a sufficiently compact substructure.  
If the host lens model where changed to something other than a SIS+shear
model substructure would still be needed if the jet is
truly straight in image~A and curved in image~B.  The new model would
also need to reproduce the positions of the center of the jet relative
to the core and the magnification ratio of the cores.  Because of this
the magnification matrix ${\bf\tilde{A}}$ could not be drastically
different. The size,
position and structure of the subclumps needed may change somewhat with the host
model, but the general conclusions would still be the same.  

\section{Implications for Dark Matter and Cosmology}
\label{sec:impl-dark-matt}

The structures responsible for the bend in image~B of B1152+199 and
the possible kink in image~A of Q0957+561 are not
terribly unusual in their mass or size.  There are dwarf galaxies and globular
clusters orbiting our galaxy that would fit the description.  The
importance lies in the likelihood of such a structure being close enough
to the image to cause observable bending.

\subsection{Estimated substructure densities}
\label{sec:estim-substr-dens}

To estimate the probability of a jet like the one in B1152+199 having an
observable bend, we will consider the bending effect of a single clump
acting by itself.  The host lens probably enhances the effect of the clump
to a small degree.  This will not change the results of this section
by a large amount and so this extra complication will be neglected.

If we consider a straight line in the source plane that passes by a
spherically symmetric lens centered at ${\bf x}_0$ with an impact
parameter of $b$ the lensing equation~(\ref{lens_eq}) can be reduced to
\begin{equation}\label{line_image}
b=\left[ r \pm \alpha_r(r) \right] \cos(\theta)~~~,~~~r>0
\end{equation}
where $r\equiv|{\bf x} - {\bf x}_0|$, $\theta$ is the corresponding
axial coordinate and $\alpha_r(r)$ is the radial deflection which is $<0$.  The
positive sign is used for $-\pi/2<\theta<\pi/2$ -- the primary image -- and
the positive sign otherwise -- the secondary image.  We are concerned
here only with the primary image; secondary images appear to form rarely
in compound lensing with the mass scales considered here
\markcite{2001ApJ...563....9M}({Metcalf} \& {Madau} 2001) and they will generally be demagnified.  

The curvature of the image can be calculated by taking derivatives of the
curve~(\ref{line_image}). At the point $\theta=0$ the curvature is ${\bf
n}(\theta=0)=\frac{1}{r^2} \left( \frac{d^2r}{d\theta^2}-r \right)
\hat{\bf x}$.  For our two models for the subclump this is
\begin{equation}\label{curvature}
{\bf n}(\theta=0) = \frac{-1}{\theta_E}\left\{
\begin{array}{lc}
\frac{1}{(x_b+\sqrt{x_b^2+4})^2}\left( \frac{3x_b^2
+8}{\sqrt{x_b^2+4}}\right)~ \hat{\bf x} & \mbox{(point mass)} \\
\frac{1}{(x_b+1)^2}~ \hat{\bf x} & \mbox{(SIS)}
\end{array} \right.
\end{equation}
where $x_b\equiv b/\theta_E$.  For the point mass $\theta_E=\sqrt{m/(\pi
D_l^2\Sigma_c)}$ and for the SIS $\theta_E=\lambda_E(\sigma)/D_l$.

A clump will not make an observable bend in a jet of length $\theta_{\rm
jet}$ if the Einstein ring radius is either too big or too small.  From
(\ref{curvature}) we see that the maximum curvature a clump can produce
is $\theta_E(z)^{-1}$.  When $\theta_E(z)$ is larger than the length of
the jet the deviation from a straight line is at
most $\sim \theta_{\rm jet}^2/8\theta_E$.  This must be larger than the
smallest measurable scale, $\theta_{\rm res}$, which is set by
either the resolution of the observations or the width of the jet.  Applying
this criterion to the curvature as a function of $b$, (\ref{curvature}),
gives an upper limit on the impact parameter.  A small clump will
influence a region of the jet of size $\sim \theta_E$.  If the smallest
scale $\theta_{\rm res}$ is of order the circumference of the Einstein
ring then its bending effects will be on too small a scale to be
observed.  These constraints are summarized as
\begin{equation}\label{inequalities}
\frac{\theta_{\rm res}}{2\pi} \simlt \theta_E \simlt
\frac{\theta^2_{\rm jet}}{8 \theta_{\rm res}} 
~~~~,~~~~ 
|{\bf n}(x_b)| \simgt \frac{8
\theta_{\rm res}}{\theta^2_{\rm jet}}
\end{equation}
The first of these inequalities can be used to find the range of
velocity dispersions or masses that could be responsible an observable
bending of the jet in B1152+199:
\begin{equation}\label{sigma_range}
6\kms \simlt \sigma \simlt 13 \kms
\end{equation}
\begin{equation}
7.1\times 10^4 h_{65}^{-1} \msun \simlt m \simlt 2.7\times 10^7 h_{65}^{-1}\msun
\end{equation}
where the values $\theta_{\rm jet}=15$~milli-arcsec and $\theta_{\rm
res}= 3$~milli-arcsec have been used.  This range is consistent with the
$\sigma$ derived in \S~\ref{sec:results-b1152+199}. 
The true ranges are probably a bit larger because of the influence of
the host lens which will increase the sensitive to smaller mass objects.
The second of the inequalities~(\ref{inequalities}) puts an upper limit
on the impact parameter $b$ as a function of $\sigma$ or $m$
through~(\ref{curvature}).  By plugging in the smallest allowed clump we
can find the largest possible impact parameter -- $b \simlt
1.6\,{\rm mas}=10 h_{65}^{-1}$~pc for the SIS and $6.9\,{\rm mas}= 28
h_{65}^{-1}$~pc for the point mass.  The clump needs to be quite well
aligned with the image.

The probability of a subclump bending the jet will be taken to be 
$p\propto \theta_{\rm jet}db$ within the allowed range of $b$.  The
probability or expected number of important clumps per jet is
\begin{equation}\label{prob}
p\simeq 2\theta_{\rm jet} \frac{c}{H_0} \int_0^{z_s}
\frac{dz}{(1+z)} \frac{D(z)^2}{E(z)} 
\int^{m_{\rm max}(z)}_{m_{\rm min}(z)} dm~
b_{\rm max}(m,z) \frac{d{\mathcal N}}{dm}(m,z)
\end{equation}
where ${\mathcal N}$ is the 3--dimensional number density of clumps and
$E(z)=\left[ \Omega_m (1+z)^3 + \Omega_R(1+z)^2 +
\Omega_\Lambda \right]^{1/2}$, $\Omega_R=1-\Omega_m-\Omega_\Lambda$.  In the case of SIS lenses $m$ can be
replaced with $\sigma$ and $b_{\rm max}(\sigma,z)$ can be found explicitly.
For the point mass case $b_{\rm max}(m)$ must be found numerically.

To get a simple estimate of the number density of clumps required we can
take them to all lie within the host lens and give them all the same
velocity dispersion.  In this case (\ref{prob}) reduces to
\begin{equation}
p\simeq 2 \theta_{\rm jet}^2  D_l^2 \,\eta(\sigma) 
\left( \sqrt{\frac{2  \theta_E(\sigma)}{\theta_{\rm res}}} - \frac{\theta_E(\sigma)}{\theta_{\rm jet}} \right)~~~~~~\mbox{(SIS)}
\end{equation}
where $\eta(\sigma)$ is the 2--dimensional number density of
clumps.  The range of allowed $\sigma$ given in (\ref{sigma_range})
gives a range $\eta(\sigma)/p \simeq 32-111~h_{65}^2\kpc^{-2}$.
This is the number density of substructures required to make the bending
commonplace.  The same exercise with point masses in the range $m=10^5 -
10^7\msun$ gives a range of $\eta(\sigma)/p \simeq
130-260~h_{65}^2\kpc^{-2}$ or $\Sigma = 1.3\times 10^7 - 2.6 \times 10^9
\msun\kpc^{-2}$ where the higher mass density is for larger mass clumps.
In units of the critical density this is $\kappa=0.005 - 0.99$.  This
value is anywhere from a few percent to more than all of the surface
density of the host lens.  The lensing effect of the host lens may reduce these
estimates by a factor of roughly $|\mu|^{-1/2}$ -- an estimate of the
eigenvalues of the magnification matrix -- which is 0.3--0.9 for the model
found in \S~\ref{sec:results-b1152+199}.

Instead of fixing the mass of the substructure we can guess at a
realistic mass function.  One expects that the number density of small
mass clumps will be proportional to the density of all matter, $\rho$,
averaged over a larger scale than the clumps being considered -- constant
Lagrangian number density.  CDM simulations and analytic estimates predict a
power--law mass function for the low mass range important here,
\begin{equation}\label{mass_function}
\frac{1}{\rho}\frac{d{\mathcal N}}{dm} = \frac{1}{M_o m_o}
\left( \frac{m}{m_o} \right)^{\alpha}
\end{equation}
where $m_o$ and $M_o$ are normalization constants.  Fitting the mass function
from $\Lambda$CDM N--body simulations to the observed velocity
distribution in the range $V_{\rm circ}=20-400\kms$ gives the relation
$\sigma \simeq 100\kms (m/3.0\times 10^{11}\msun)^{1/3}$.  For SIS
substructures this relation is used to convert (\ref{mass_function})
into a distribution of velocity dispersions where it is extrapolate
below $V_{\rm circ}=20\kms$.  
In $\Lambda$CDM simulations the dark matter clumps
have $\alpha \simeq -1.91$ and $M_o=4.8\times 10^{12} h^{-1}\msun$ for
$m_o=3.0\times 10^{11}\msun$ \markcite{1999ApJ...522...82K}({Klypin} {et~al.} 1999).  The exponent
for the $\sigma$ distribution is $\alpha_\sigma=-3.73$ in this case.
This distribution fits the observed distribution of dwarf galaxies near 
$\sigma = 50-100 \kms$ above which the contribution to (\ref{prob})
is small.

Using the full range of masses in~(\ref{sigma_range}) and keeping all
the subclumps at the redshift of the host lens results in a probability of
$p \simeq 3.2 \,\kappa$ where $\kappa$ is the surface density of the host
lens,  $\kappa=0.35$ and $0.85$ for the model in \S~\ref{sec:results-b1152+199}.
Figure~\ref{fig:prob_fsub} shows $p$ and the fraction of the halo mass density
contained in substructure as a function of a lower mass cutoff in the
mass function.  The smaller mass clumps contribute most of the
probability, but little of the mass density.  This mass fraction is a
lower limit in that if the internal structure of the subclumps is less
centrally concentrated it will require more mass to reach the same
probability.  For SIS substructures that are not tidally truncated $p=
1.9\times 10^{-3}\,\kappa$.  To increase this probability by a factor of
ten would require the entire mass density of the host lens to be
composed of SISs in the range~(\ref{sigma_range}).  Any tidal truncation
will reduce SIS substructures' lensing effect. 

Objects that are not in the host galaxy, but happen to lie near the line
of sight could also cause bending of the jet.  To estimate this 
contribution we integrate~(\ref{prob}) with the mass
function~(\ref{mass_function}) assuming that $\rho$ along the line of
sight is given by the average density of the universe.  For SIS
structure $p=1.3\times 10^{-4}$ and for point masses with the
same mass function $p=0.65$.  This extra--galactic population is only an
important contribution to the probability if the clumps are very compact
in which case it is comparable to the contribution from substructures
inside the lens.

The CDM model does seem capable of accounting for the bent jets, provided
DM halos are relatively compact.  If the radius is small compared to the
Einstein radius of a point mass of the same mass ($r\simlt \theta_E = 11
(m/10^6\msun)^{1/2} h_{65}^{1/2} \mbox{ pc}$) less than $\sim 10\%$ of
the mass need be in substructure.  However, any less concentrated clumps will 
require more total mass.  The SISs require much more mass.  The Navarro,
Frenk \& White (NFW) profile \markcite{1997ApJ...490..493N}({Navarro}, {Frenk}, \&  {White} 1997), $\rho(r) =
\rho_c r_s^3 r^{-1}(r_s+r)^{-2}$, is believed to be more realistic for
pure dark matter halos.  If $r_s$ is small compared to the above limit
and a large fraction of the mass is within this radius then the mass
fraction might get down to the levels shown in figure~\ref{fig:prob_fsub}.
The scale length according to the standard structure formation scenario
is $r_s= 2.17\times 10^3 {\rm c}^{-1}
h_{65}^{-2/3}(m_{200}/10^6\msun)^{1/3}\mbox{ pc}$ where c is the 
concentration and $m_{200}$ is the virial mass.  If the concentration is
100 or larger then the core is compact enough, but in this case the mass within
$r_s$ is less than 10\% of $m_{200}$.  In addition, $c\simeq 100$ is
a bit high for a straightforward extrapolation of the simulations
\markcite{astro-ph/9908159}({Bullock} {et~al.} 2001) -- no
simulation has been done with a resolution high enough to resolve these
mass scales.  To achieve the same probability for
bending the jet, it seems that any realistic CDM model will require
significantly more mass -- at least before tidal stripping occurs -- to
be in small scale structure than is required in the point mass model used
here. 

Also, the survival of substructure in the host lens is a
complicated issue. Clumps with $m \simgt 10^7\msun$ are not likely
to survive within the inner few kpc because they lose orbital energy
to dynamical friction and fall into the center of the galaxy where they
are destroyed by tides.  This upper mass cutoff can significantly change
the local fraction of mass in substructures while not affecting the lensing
probability greatly.

\subsection{Contribution from known structures}
\label{sec:contr-from-known}

There are about 40 known dwarf galaxies in the Local Group 
\markcite{1998ARA&A..36..435M,1999ApJ...522...82K}({Mateo} 1998; {Klypin} {et~al.} 1999).  Most of these are
within $\sim 300\kpc$ of either the MW or M31.
About twenty eight of these have circular velocities above $10\kms$.
This gives an estimated surface number density of $\sim 3.5\times
10^{-5}\kpc^{-2}$ if they were uniformly distributed in this volume.
There are about 200 globular clusters in the MW with masses of
$10^4 - 10^6 \msun$ making their number density an order of magnitude
larger.  The concentration of dwarfs and globular clusters toward the center of
the galaxy and observational incompleteness might increase this estimate
by a factor of several, but nowhere near enough 
to reach the required number densities derived in the previous section.

\begin{figure}[t]
\centering\epsfig{figure=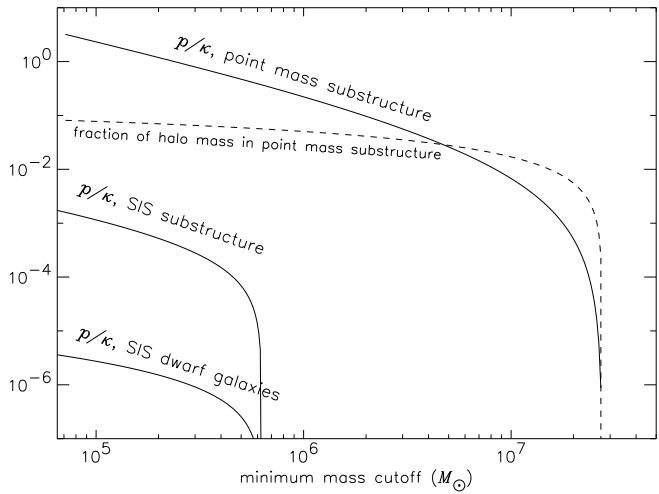,height=2.75in}
\caption[qiw]{\footnotesize The probability of substructures causing an
observable bend in a radio jet like the one in B1152+199 assuming the
distribution of substructures described in the text.  For the SIS
substructures the velocity dispersion is converted to mass by
$m=3.0\times 10^{11}\msun (\sigma/100\kms)^3$.  The fraction of the host
halo surface density contained in point mass substructure is also plotted.
The host lens surface density is $\kappa=0.6-1.4$ for the model of B1152+199
discussed in \S~\ref{sec:results-b1152+199}.
}
\label{fig:prob_fsub}
\end{figure}

Another way of estimating the contribution from dwarf galaxies is to use
the mass function~(\ref{mass_function}) converted to velocity
dispersion.  For the observed galaxies within $200h^{-1}$~kpc of the
MW and M31 $\alpha_\sigma = -2.35\pm 0.4$ and $m_o\simeq
M(<200)/6.32$ for $\sigma_o=10\kms$ where $M(<200)$ is the total mass
within $200h^{-1}$~kpc \markcite{1999ApJ...522...82K}({Klypin} {et~al.} 1999).  We will use
$M(<200)=10^{12}\msun$.  With SIS dwarf galaxies this velocity
distribution gives a probability for bending the jet of $p=3.9\times
10^{-6}\, \kappa$ if the dwarfs are in the host 
lens.  Figure~\ref{fig:prob_fsub} shows $p$ as a function of a lower
$\sigma$ cutoff which is converted into mass by $\sigma = 100\kms
(m/3.0\times 10^{11}\msun)^{1/3}$.   If the same velocity
distribution is used for the entire line of sight at the average mass
density, $p=2.7\times 10^{-7}$.  Dwarf galaxies are not compact enough to
be considered point mass lenses, but by treating them as point masses 
we can get an (probably greatly inflated) upper limit on the probability.  In
this case $p=8.9\times 10^{-5} \, \kappa$.

Known types of substructure within the host lens are inadequate to
explain B1152+199.  If the structures in the lens and in intergalactic
space are similar in number and central density to those observed in the
local group of galaxies they fall short of the estimates derived in
\S~\ref{sec:estim-substr-dens} by at least a factor of $10^5$.

\section{Discussion}
\label{sec:discussion}

These observations have important consequences for the Warm Dark Matter
(WDM) model. The standard WDM model is engineered to reproduces the dwarf
galaxy distribution under the assumption that a galaxy forms in every
small halo.  It was shown in \S~\ref{sec:contr-from-known} that the number
density of dwarf galaxies is extremely unlikely to have produced the
observed bent radio jets.  The standard WDM model is thus ruled out if
the bend in B1152+199 is real.  A more accurate lower limit on the DM
particle mass will require more observations and more simulations of
small scale structure formation in these models.

Higher resolution observations of B1152+199 are possible.  These would
make certain that the jet in image~B is indeed bent and improve the
constraints on the substructure mass.  Also interesting would be
high resolution images of other multiply imaged jets.  In the present sample of
three all appear to show some evidence of bending.  A moderately larger
sample would greatly increase the power of this method to probe
structure on small scales.

It has been found here that a significantly larger number of small scale 
objects are needed if the observations of B1152+199 are to be simply
interpreted.  Structures as diffuse as SIS are disfavored both by direct
modeling of B1152+199 and on statistical grounds.  If the structures are compact
(on the scale of their own Einstein radius) and small in mass ($\simlt
10^7\msun$) they need not 
contain a large fraction of the mass in the universe.  However, such 
concentrated halos come about in the CDM model only through the tidal
stripping of halos that originally contained $\sim 10$ times more mass.
This means that in intergalactic space these clumps would contain a
large fraction of the mass, perhaps most of it.

\acknowledgments
I would like to thank P. Madau, M. Magliocchetti and H. Zhao for useful
discussions and comments.  Special thanks to D. Rusin for bringing
the case of B1152+199 to my attention and providing the VLBI maps.


\begin{thebibliography}{}

\bibitem[{Barkana}, {Leh{\'a}r}, {Falco}, {Grogin},  {Keeton}, \& {Shapiro} 1999]{1999ApJ...520..479B}
{Barkana}, R., {Leh{\'a}r}, J., {Falco}, E.~E., {Grogin}, N.~A., {Keeton},  C.~R., \& {Shapiro}, I.~I. 1999, \apj, 520, 479

\bibitem[{Blandford} \& {Jaroszynski} 1981]{1981ApJ...246....1B}
{Blandford}, R.~D. \& {Jaroszynski}, M. 1981, \apj, 246, 1

\bibitem[{Bode}, {Ostriker}, \&  {Turok} 2001]{2001ApJ...556...93B}
{Bode}, P., {Ostriker}, J.~P., \& {Turok}, N. 2001, \apj, 556, 93

\bibitem[Brada\v{c}, Schneider, Steinmetz, Lombardi,  King, \& Porcas 2002]{astro-ph/0112038}
Brada\v{c}, M., Schneider, P., Steinmetz, M., Lombardi, M., King, L., \&  Porcas, R. 2002, preprint, astro-ph/0112038

\bibitem[{Bullock}, {Kolatt}, {Sigad}, {Somerville},  {Kravtsov}, {Klypin}, {Primack}, \& {Dekel} 2001]{astro-ph/9908159}
{Bullock}, J.~S., {Kolatt}, T.~S., {Sigad}, Y., {Somerville}, R.~S.,  {Kravtsov}, A.~V., {Klypin}, A.~A., {Primack}, J.~R., \& {Dekel}, A. 2001,  \mnras, 321, 559

\bibitem[{Bullock}, {Kravtsov}, \&  {Weinberg} 2000]{2000ApJ...539..517B}
{Bullock}, J.~S., {Kravtsov}, A.~V., \& {Weinberg}, D.~H. 2000, \apj, 539, 517

\bibitem[{Chiba} 2002]{2002ApJ...565...17C}
{Chiba}, M. 2002, \apj, 565, 17

\bibitem[Dalal \& Kochanek 2002]{Dalal2002}
Dalal, N. \& Kochanek, C. 2002, preprint, astro-ph/0111456

\bibitem[{Garrett}, {Calder}, {Porcas}, {King},  {Walsh}, \& {Wilkinson} 1994]{1994MNRAS.270..457G}
{Garrett}, M.~A., {Calder}, R.~J., {Porcas}, R.~W., {King}, L.~J., {Walsh}, D.,  \& {Wilkinson}, P.~N. 1994, \mnras, 270, 457

\bibitem[{Irwin}, {Webster}, {Hewett}, {Corrigan}, \&  {Jedrzejewski} 1989]{1989AJ.....98.1989I}
{Irwin}, M.~J., {Webster}, R.~L., {Hewett}, P.~C., {Corrigan}, R.~T., \&  {Jedrzejewski}, R.~I. 1989, \aj, 98, 1989

\bibitem[Kamionkowski \& Liddle 2000]{2000PRL.Kamionkowski}
Kamionkowski, M. \& Liddle, A. 2000, PRL, 84, 4525

\bibitem[Keeton 2002]{Keeton2002}
Keeton, C. 2002, preprint, astro-ph/0112350

\bibitem[{Kemball}, {Patnaik}, \&  {Porcas} 2001]{2001ApJ...562..649K}
{Kemball}, A.~J., {Patnaik}, A.~R., \& {Porcas}, R.~W. 2001, \apj, 562, 649

\bibitem[{King}, {Browne}, {Muxlow}, {Narasimha},  {Patnaik}, {Porcas}, \& {Wilkinson} 1997]{1997MNRAS.289..450K}
{King}, L.~J., {Browne}, I.~W.~A., {Muxlow}, T.~W.~B., {Narasimha}, D.,  {Patnaik}, A.~R., {Porcas}, R.~W., \& {Wilkinson}, P.~N. 1997, \mnras, 289,  450

\bibitem[{Klypin}, {Kravtsov}, {Valenzuela}, \&  {Prada} 1999]{1999ApJ...522...82K}
{Klypin}, A., {Kravtsov}, A.~V., {Valenzuela}, O., \& {Prada}, F. 1999, \apj,  522, 82

\bibitem[{Koopmans}, {Bruyn}, {Marlow}, {Jackson},  {Blandford}, {Browne}, {Fassnacht}, {Myers}, {Pearson}, {Readhead},  {Wilkinson}, \& {Womble} 1999]{1999MNRAS.303..727K}
{Koopmans}, L.~V.~E., {Bruyn}, A.~G.~D., {Marlow}, D.~R., {Jackson}, N.,  {Blandford}, R.~D., {Browne}, I.~W.~A., {Fassnacht}, C.~D., {Myers}, S.~T., {et al.}, 1999, \mnras, 303, 727

\bibitem[{Mao} \& {Schneider} 1998]{1998MNRAS.295..587M}
{Mao}, S. \& {Schneider}, P. 1998, \mnras, 295, 587

\bibitem[{Marlow}, {Rusin}, {Norbury}, {Jackson},  {Browne}, {Wilkinson}, {Fassnacht}, {Myers}, {Koopmans}, {Blandford},  {Pearson}, {Readhead}, \& {de Bruyn} 2001]{2001AJ....121..619M}
{Marlow}, D.~R., {Rusin}, D., {Norbury}, M., {Jackson}, N., {Browne}, I.~W.~A.,  {Wilkinson}, P.~N., {Fassnacht}, C.~D., {Myers}, S.~T., {et al.}, 2001, \aj, 121, 619

\bibitem[{Mateo} 1998]{1998ARA&A..36..435M}
{Mateo}, M.~L. 1998, \araa, 36, 435

\bibitem[Metcalf 2001]{astro-ph/0109347}
Metcalf, R. 2001, in Where is the Matter?, ed. L.~Tresse \& M.~Treyer  (astro-ph/0109347)

\bibitem[{Metcalf} \& {Madau} 2001]{2001ApJ...563....9M}
{Metcalf}, R.~B. \& {Madau}, P. 2001, \apj, 563, 9

\bibitem[{Metcalf} \& {Zhao} 2002]{2002ApJ...567L...5M}
{Metcalf}, R.~B. \& {Zhao}, H. 2002, \apjl, 567, L5

\bibitem[{Moore}, {Ghigna}, {Governato}, {Lake},  {Quinn}, {Stadel}, \& {Tozzi} 1999]{1999ApJ...524L..19M}
{Moore}, B., {Ghigna}, S., {Governato}, F., {Lake}, G., {Quinn}, T., {Stadel},  J., \& {Tozzi}, P. 1999, \apjl, 524, L19

\bibitem[{Myers}, {Rusin}, {Fassnacht}, {Blandford},  {Pearson}, {Readhead}, {Jackson}, {Browne}, {Marlow}, {Wilkinson},  {Koopmans}, \& {de Bruyn} 1999]{1999AJ....117.2565M}
{Myers}, S.~T., {Rusin}, D., {Fassnacht}, C.~D., {Blandford}, R.~D., {Pearson},  T.~J., {Readhead}, A.~C.~S., {Jackson}, N., {Browne}, I.~W.~A., {et al.}, 1999,  \aj, 117, 2565

\bibitem[{Navarro}, {Frenk}, \&  {White} 1997]{1997ApJ...490..493N}
{Navarro}, J.~F., {Frenk}, C.~S., \& {White}, S. D.~M. 1997, \apj, 490, 493

\bibitem[{Ros}, {Guirado}, {Marcaide}, {P{\'  e}rez-Torres}, {Falco}, {Mu{\~ n}oz}, {Alberdi}, \&  {Lara} 2000]{2000A&A...362..845R}
{Ros}, E., {Guirado}, J.~C., {Marcaide}, J.~M., {P{\' e}rez-Torres}, M.~A.,  {Falco}, E.~E., {Mu{\~ n}oz}, J.~A., {Alberdi}, A., \& {Lara}, L. 2000, \aap,  362, 845

\bibitem[{Rusin}, {Marlow}, {Norbury}, {Browne},  {Jackson}, {Wilkinson}, {Fassnacht}, {Myers}, {Koopmans}, {Blandford},  {Pearson}, {Readhead}, \& {de Bruyn} 2001]{2001AJ....122..591R}
{Rusin}, D., {Marlow}, D.~R., {Norbury}, M., {Browne}, I.~W.~A., {Jackson}, N.,  {Wilkinson}, P.~N., {Fassnacht}, C.~D., {Myers}, S.~T., {et al.}, 2001, \aj, 122, 591

\bibitem[{Rusin}, {Norbury}, {Biggs}, {Marlow},  {Jackson}, {Browne}, {Wilkinson}, \& {Myers} 2002]{2002MNRAS.330..205R}
{Rusin}, D., {Norbury}, M., {Biggs}, A.~D., {Marlow}, D.~R., {Jackson}, N.~J.,  {Browne}, I.~W.~A., {Wilkinson}, P.~N., \& {Myers}, S.~T. 2002, \mnras, 330,  205

\bibitem[Somerville 2002]{astro-ph/0107507}
Somerville, R.~S. 2002, preprint, astro-ph/0107507

\bibitem[{Spergel} \& {Steinhardt} 2000]{2000PhRvL..84.3760S}
{Spergel}, D.~N. \& {Steinhardt}, P.~J. 2000, Physical Review Letters, 84, 3760

\bibitem[{van den Bosch}, {Robertson},  {Dalcanton}, \& {de Blok} 2000]{2000AJ....119.1579V}
{van den Bosch}, F.~C., {Robertson}, B.~E., {Dalcanton}, J.~J., \& {de Blok},  W.~J.~G. 2000, \aj, 119, 1579

\bibitem[{Walsh}, {Carswell}, \&  {Weymann} 1979]{1979Natur.279..381W}
{Walsh}, D., {Carswell}, R.~F., \& {Weymann}, R.~J. 1979, \nat, 279, 381

\bibitem[{Wambsganss} \& {Paczynski} 1992]{1992ApJ...397L...1W}
{Wambsganss}, J. \& {Paczynski}, B. 1992, \apjl, 397, L1

\bibitem[{Witt}, {Mao}, \& {Schechter} 1995]{witt:95}
{Witt}, H.~J., {Mao}, S., \& {Schechter}, P.~L. 1995, \apj, 443, 18

\bibitem[{Wo{\' z}niak}, {Alard}, {Udalski},  {Szyma{\' n}ski}, {Kubiak}, {Pietrzy{\' n}ski}, \& {Zebru{\'  n}} 2000]{2000ApJ...529...88W}
{Wo{\' z}niak}, P.~R., {Alard}, C., {Udalski}, A., {Szyma{\' n}ski}, M.,  {Kubiak}, M., {Pietrzy{\' n}ski}, G., \& {Zebru{\' n}}, K. 2000, \apj, 529,  88

\bibitem[{Xanthopoulos}, {Norbury}, {Karidis},  {Jackson}, {Browne}, {Wilkinson}, {Porcas}, {Patnaik}, \&  {Gabuzda} 2000]{2000evn..proc...49X}
{Xanthopoulos}, E., {Norbury}, M., {Karidis}, A., {Jackson}, N.~J., {Browne},  I.~W.~A., {Wilkinson}, P.~N., {Porcas}, A.~R., {Patnaik}, A.~R., {et al.}, 2000, in EVN Symposium 2000, Proceedings of the 5th european  VLBI Network Symposium held at Chalmers University of Technology, Gothenburg,  Sweden, June 29 - July 1, 2000, Eds.: J.E. Conway, A.G. Polatidis, R.S. Booth  and Y.M. Pihlstr{\" o}m, published Onsala Space Observatory, p. 49, 49+

\end{thebibliography}
\end{document}